\documentclass[12pt,33, final]{iopart}

\expandafter\let\csname equation*\endcsname\relax
\expandafter\let\csname endequation*\endcsname\relax

\usepackage{color}
\usepackage{graphicx}
\usepackage{cancel}
\usepackage{amsfonts}
\usepackage{amssymb}
\usepackage{amsthm}
\usepackage{yhmath}
\usepackage{amsmath}
\usepackage{svg}
\usepackage{url}
\usepackage{hyperref}
\usepackage{float}
\usepackage{multirow}  
\usepackage{dcolumn}   
\usepackage{bm}        
\usepackage{enumerate}
\usepackage{stmaryrd}
\hypersetup{
	colorlinks,
	linkcolor={green!50!black},
	urlcolor={blue!80!black},
	citecolor   = {red!50!black} 
}

\def \bO       {\mbox{\boldmath$\Omega$}}

\usepackage{iopams}  
\begin{document}

\title[{\footnotesize{Intrinsic AHE, SHE and VHE in ex-so-tic vdW structures}}]{Intrinsic anomalous, spin and valley Hall effects in ex-so-tic van-der-Waals structures} 

\author{I. Wojciechowska and A. Dyrdał}

\address{Faculty of Physics, ISQI, Adam Mickiewicz University in Poznań, Poland\\ ul. Uniwersytetu Poznańskiego 2, 61-614 Poznań}
\ead{adyrdal@amu.edu.pl}
\vspace{10pt}

\begin{abstract}
We consider the anomalous, spin, valley, and valley spin Hall effects in a pristine ex-so-tic graphene-based van-der-Waals (vdW) heterostructure consisting of a bilayer graphene (BLG) between semiconducting  van-der-Waals material with  strong SOC (e.g., WS$_2$) and ferromagnetic and insulating vdW material (e.g. Cr$_2$Ge$_2$Te$_6$). Reducing the effective Hamiltonian derived by Zollner et al [Phys. Rev. Lett. 125(19), 196402 (2020)] to low-energy states, and using the Green function formalism, we derived analytical  results for the Hall conductivities as a function of the Fermi level and gate voltage. Depending on these parameters, we found quantized valley conductivity. 
\end{abstract}

%
\vspace{2pc}

\section{Introduction}

Two-dimensional (2D) van-der-Waals (vdW) materials, being promising materials for further development of spintronics and a real step towards further miniaturization of electronic devices~\cite{Jin_Nanoscale_Rev_2023,MRS2020}, focus nowadays an enormous attention. The most fascinating aspect of van-der-Waals structures is the possibility of designing electronic, magnetic and topological properties on demand due to a simple consequence of stacking selected 2D van-der-Waals crystals (having specific physical properties) and proximity effects that emerge in such stacks~\cite{Gaim_Grigorieva_2013}. 
Accordingly, van-der-Waals hybrid structures, created by stacking 2D crystals that reveal a wide range of physical properties and phases  (e.g., insulators, semiconductors, metals, superconductors, magnetics, ferroelectrics, etc.~\cite{Basov_NatMat2017,Tokura_Kawasaki_Nagaosa_NatPhys2017,Giustino_2020,Weber_2024,Wang_Rev_Genome_ACS2022}), constitute a unique class of materials with a combination of different physical properties, that can be tuned not only by external fields and forces but also due to mutual coupling between different phases of matter within the structure. This is the case, for example, in van-der-Waals multiferroics~\cite{Cui_npj2DMatandAppl_2018,Zhang_NatRevPhys_2023}, that can be obtained by stacking two van-der-Waals crystals: ferroelectric and ferromagnetic ones.

According to the main paradigm of spin-electronics, which assumes the usage of electron spin on equal footing with its charge, the all-electrical control of the spin degree of freedom and search for additional degrees of freedom that can be coupled to the spin are of special interest~\cite{Sinova_SpintronicsTango2012}. Naturally, the spin-orbit-driven transport phenomena, such as current-induced spin polarization, anomalous and spin Hall effects and their quantum counterparts~\cite{Nagaosa_AHE_RevModPhys2010,Sinova_RevModPhys2015,Jin_Nanoscale_Rev_2023,Jungwirth_Olejnik_NatMat_2012} have become a hallmark of modern spin electronics. Importantly, van-der-Waals hybrid structures with their additional degrees of freedom, such as a valley or pseudospin (related to the sublattices and other orbital properties of vdW structures), seem to be a perfect platform for designing a new generation of spintronic devices~\cite{Giustino_2020,Weber_2024,Pesin_MacDonald_NatMat2012,NatRevMat_valleytronics_2016,Liu_NanoResearch2019,Valenzuela_Roche_Nature2022}. 
Among the van-der-Waals hybrid structures, the stacks containing graphene are under great attention. In this particular case, one can design structures that explore the unique and high-quality electronic properties of graphene, enriched with additional physical properties revealed as a consequence of the proximity effects due to its contact with adjacent layers (see for example~\cite{Gmitra_PRB_GGR/TMDC2015,Yang_2016,Zollner_PRB2016,Hallal_2017,Dyrdal2017,Zollner_exsotic,Zollner_CGT/BL/CGT_TMDC/BL/TMDC,Avsar_NatComm2014,Rezende_PhysRevLett2015,Leutenantsmeyer_2017,Roche_Casanova_NanoLett-2019,Valenzuela_Roche-2020,Inglot_PRB2021,Vila2021,Pezo_2022} and reviews~\cite{Abergel2010,Ferrari_Roche_Roadmap_Nanoscale2015,Roche_NuovoCimento2016,Fabian_Roche_NatNano2021,Kurebayashi2022,AFerreira_EncyCondMatt2024}). 

\begin{figure}[t!] 
\raggedright 
\includegraphics[width=0.99\textwidth]{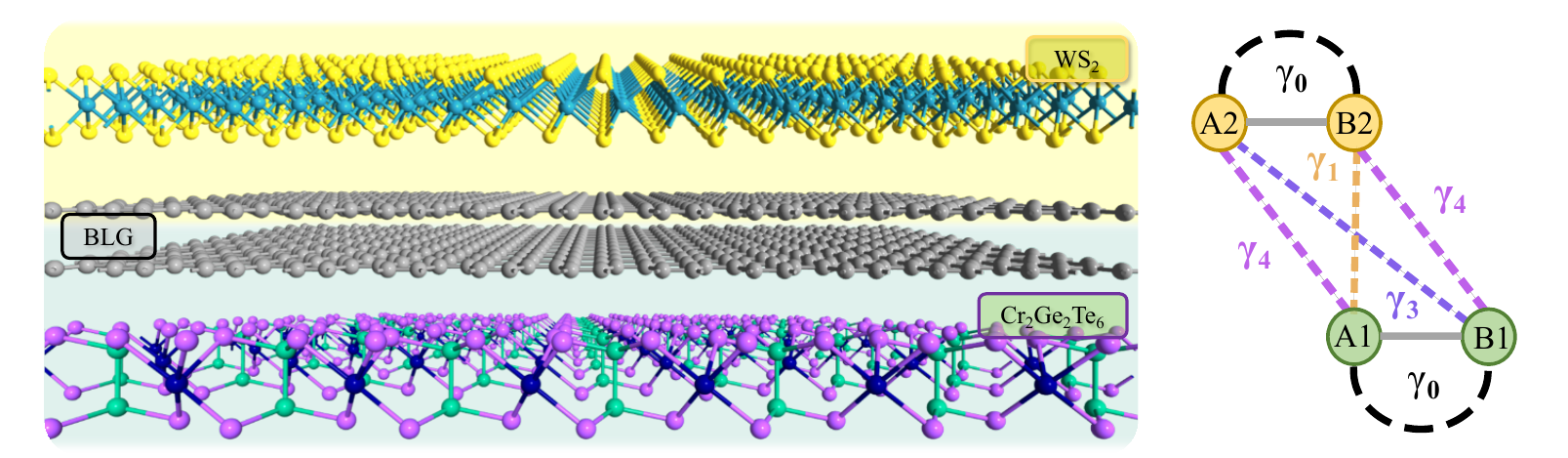}
\caption{Ex-so-tic van-der Waals structure (on the left) consisting of a bilayer graphene (BLG) between other 2D vdW crystals: semiconducting with strong SOC (e.g., transition metal dichalcogenide such as  WS$_2$) and ferromagnetic insulator such as Cr$_2$Ge$_2$Te$_6$. The intralayer hopping parameter $\gamma_{0}$ and interlayer hopping ones $\gamma_{0,1,3,4}$ taken into account in the considered Hamiltonian of BLG are indicated on the right side. Here A1,2 (B1,2) correspond to the graphene top (1) or bottom (2) layer, and graphene sublattice A or B.}
\label{fig:fig1}
\end{figure}

Here we consider the  Hall effects in a pristine ex-so-tic graphene-based van-der-Waals heterostructure~\cite{Zollner_exsotic}. In other words, we restrict our considerations to transport properties in the clean limit, i.e., we do not consider the effect of spin-orbital scatterers, that could lead to an extrinsic contribution to the Hall conductivities due to the skew-scattering or side-jump mechanisms~\cite{Nagaosa_AHE_RevModPhys2010}.  The name 'ex-so-tic structure' originates from the building blocks in the structure under consideration. Namely, a bilayer of graphene  (BLG) is deposited on a magnetic 2D insulator, e.g., on ${\mathrm{Cr_{2}Ge_{2}Te_{6}}}$ (CGT), that creates a strong proximity-induced exchange field in the bottom (1st) layer of BLG and only weakly affects the electronic properties of the top (2nd) layer. Additionally, from the top, the BLG is covered by a semiconducting 2D transition metal dichalcogenide (TMDC), i.e. ${\mathrm{WS_2}}$, responsible for a strong proximity-induced spin-orbit coupling in the top layer of BLG and a weak one in the bottom layer. The structure is presented in Fig~\ref{fig:fig1}. Interestingly, the electronic band structure is highly tunable by an external gate voltage that dramatically changes the impact of the spin-orbit and exchange interactions on the electronic conduction and valence bands (for details, see~\cite{Zollner_exsotic}).

The paper is organized as follows. In Sec.~\ref{sec:Model} we introduce the  $\bf{k}\cdot\bf{p}$ Hamiltonian provided in ~\cite{Zollner_exsotic} and derive a reduced Hamiltonian that describes the low-energy electronic spectrum around the K and K' points. The reduced Hamiltonian is used in the following sections to derive all the analytical results.  
Then, in Sec.~\ref{sec:Method} we present the formalism, which we use in Sec.~\ref{sec:Results}  to calculate the intrinsic components of the anomalous, spin Hall, valley, and spin-valley Hall effects. The general summary and conclusions are provided in Sec.~\ref{sec:Conclusions}.

\section{Model}
\label{sec:Model}
We consider ex-so-tic graphene-based van-der-Waals heterostructure, i.e. the system consisting of a bilayer of graphene (BLG) deposited on the magnetic 2D insulator, ${\mathrm{Cr_{2}Ge_{2}Te_{6}}}$ (CGT), that creates a strong proximity-induced exchange field in the bottom layer of BLG, and covered by a semiconducting 2D TMDC, 
${\mathrm{WS_2}}$, responsible for a strong proximity spin-orbit coupling in the top layer of BLG. The structure is presented schematically in Fig~\ref{fig:fig1}.

\subsection{$\mathbf{k}\cdot\mathbf{p}$ Hamiltonian}

The effective ${\mathbf{k} \cdot \mathbf{p}}$ Hamiltonian describing CGT/BLG/TMDC structure has been derived in Ref.~\cite{Zollner_exsotic} based on DFT modelling and symmetry considerations. This Hamiltonian describes electronic states in the vicinity of K and K' points (indexed by $\tau = \pm1$) of the Brillouin zone~\cite{Zollner_exsotic,KonschuhGmitra2012,KochanFabian2017}:
\begin{equation}
\label{eq:H_kp}
 \hat{H}^{\tau} = \hat{H}_{\scriptscriptstyle{ORB}}^{\tau} + \hat{H}_{\scriptscriptstyle{SOC}}^{\tau} + \hat{H}_{\scriptscriptstyle{R}}^{\tau} + \hat{H}_{\scriptscriptstyle{EX}}.    
\end{equation}

\noindent The first term of the above Hamiltonian describes the orbital physics of the structure and takes the form~\cite{Zollner_exsotic}:
\begin{equation}
\begin{split}
\hat{H}_{\scriptscriptstyle{ORB}}^{\tau} & =  - \dfrac{\sqrt{3}\gamma_{0}a}{2} \hat{\mu}_{0} \otimes (\tau k_{x} \hat{\sigma}_{x} + k_{y}\hat{\sigma}_{y}) \otimes \hat{s}_{0} 
           + \dfrac{\gamma_{1}}{2} (\hat{\mu}_{x} \otimes \hat{\sigma}_{x} - \hat{\mu}_{y} \otimes \hat{\sigma}_{y}) \otimes \hat{s}_{0}\\
        &  - \dfrac{\sqrt{3}\gamma_{3}a}{4} \hat{\mu}_{x} \otimes (\tau k_{x} \hat{\sigma}_{x} - k_{y}\hat{\sigma}_{y}) \otimes \hat{s}_{0} 
        - \dfrac{\sqrt{3}\gamma_{3}a}{4} \hat{\mu}_{y} \otimes (\tau k_{x} \hat{\sigma}_{y} + k_{y}\hat{\sigma}_{x}) \otimes \hat{s}_{0} 
        \\
         & - \dfrac{\sqrt{3}\gamma_{4}a}{2} (\tau k_{x} \hat{\mu}_{x} - k_{y}\hat{\mu}_{y}) \otimes \hat{\sigma}_{0} \otimes \hat{s}_{0} \\
         &
         + V \hat{\mu}_{z} \otimes \hat{\sigma}_{0} \otimes \hat{s}_{0} 
         + \Delta ( \hat{\mu}_{+} \otimes \hat{\sigma}_{+} + \hat{\mu}_{-} \otimes \hat{\sigma}_{-} ) \otimes \hat{s}_{0}, \\
\end{split}
\end{equation}
where $\gamma_{0,1,3,4}$ define the intralayer electron hopping between nearest neighbors, as well as interlayer hoppings between nearest and next nearest sites, as indicated in Fig.\ref{fig:fig1}, $a$ is the lattice constant of graphene, $V$ describes an effect of gate voltage (transverse displacement field), and $\Delta$ is the so-called orbital gap, being a consequence of the asymmetry in the energy shift of the bonding and antibonding states. \newline
The next term describes the intrinsic spin-orbital proximity effect and has the following form~\cite{Zollner_exsotic}:
 \begin{equation}
 \begin{split}
     \hat{H}^{\tau}_{\scriptscriptstyle{SOC}} & = \hat{\mu}_{+} \otimes \tau (\lambda_{\scriptscriptstyle{I}}^{\scriptscriptstyle{\scriptscriptstyle{A1}}}\hat{\sigma}_{+} + \lambda_{\scriptscriptstyle{I}}^{\scriptscriptstyle{B1}}\hat{\sigma}_{-}) \otimes \hat{s}_{z}
             - \hat{\mu}_{-} \otimes \tau (\lambda_{\scriptscriptstyle{I}}^{\scriptscriptstyle{A2}}\hat{\sigma}_{+} + \lambda_{\scriptscriptstyle{I}}^{\scriptscriptstyle{B2}}\hat{\sigma}_{-}) \otimes \hat{s}_{z},
 \end{split}    
 \end{equation}
with the spin-orbit coupling constant $\lambda_{I}^{Xn}$, where $X = \{A,B\}$ and $n=\{1,2\}$ indicate sublattice ($A$ or $B$) in the top (1) or bottom (2) layer, $\hat{\sigma}_{\pm} = (\hat{\sigma}_{z} \pm \hat{\sigma}_{0})/2$, and $\hat{\mu}_{\pm} =  (\hat{\mu}_{z} \pm \hat{\mu}_{0})/2$. \newline
The Rashba Hamiltonian for BLG reads~\cite{Zollner_exsotic}:
 \begin{equation}
 \begin{split}
    \hat{H}^{\tau}_{\scriptscriptstyle{R}} & = \dfrac{1}{2} (\lambda_{\scriptscriptstyle{IR}}\hat{\mu}_{z} + 2\lambda_{\scriptscriptstyle{BR}}\hat{\mu}_{0}) \otimes (\tau \hat{\sigma}_{x} \otimes \hat{s}_{y} - \hat{\sigma}_{y} \otimes \hat{s}_{x}),
 \end{split}    
 \end{equation}
where $\lambda_{\scriptscriptstyle{IR}}$ describes the strength of the  so-called intrinsic Rashba spin-orbit coupling (or Dresselhaus-like SOC), originating from a local bulk-inversion-asymmetry due to the contact with adjacent layers; and $\lambda_{\scriptscriptstyle{BR}}$ is the Bychkov-Rashba coupling constant due to the global space symmetry breaking~\cite{KonschuhGmitra2012}.

\noindent The magnetic proximity effect responsible for the exchange interaction in BLG as a result of interaction with an adjacent magnetic layer with out-of-plane anisotropy (i.e., magnetization oriented in the z-direction) is described by the following term~\cite{Zollner_exsotic}:   
\begin{equation}
\hat{H}_{\scriptscriptstyle{EX}}  =  \hat{\mu}_{+} \otimes (-\lambda_{\scriptscriptstyle{EX}}^{\scriptscriptstyle{A1}}\hat{\sigma}_{+} + \lambda_{\scriptscriptstyle{EX}}^{\scriptscriptstyle{B1}}\hat{\sigma}_{-}) \otimes \hat{s}_{z}
              - \Hat{\mu}_{-} \otimes  (-\lambda_{\scriptscriptstyle{EX}}^{\scriptscriptstyle{A2}}\hat{\sigma}_{+} + \lambda_{\scriptscriptstyle{EX}}^{\scriptscriptstyle{B2}}\hat{\sigma}_{-}) \otimes \hat{s}_{z}\hspace{1cm}
\end{equation}
where $\lambda_{\scriptscriptstyle{EX}}^{\scriptscriptstyle{Xn}}$ ($X = \{A,B\}$, $n=\{1,2\}$) is the parameter describing the strength of proximity exchange coupling in the sublattice X of the n-th layer.
\begin{figure}[t!] \raggedright 
\includegraphics[width=0.55\textwidth]{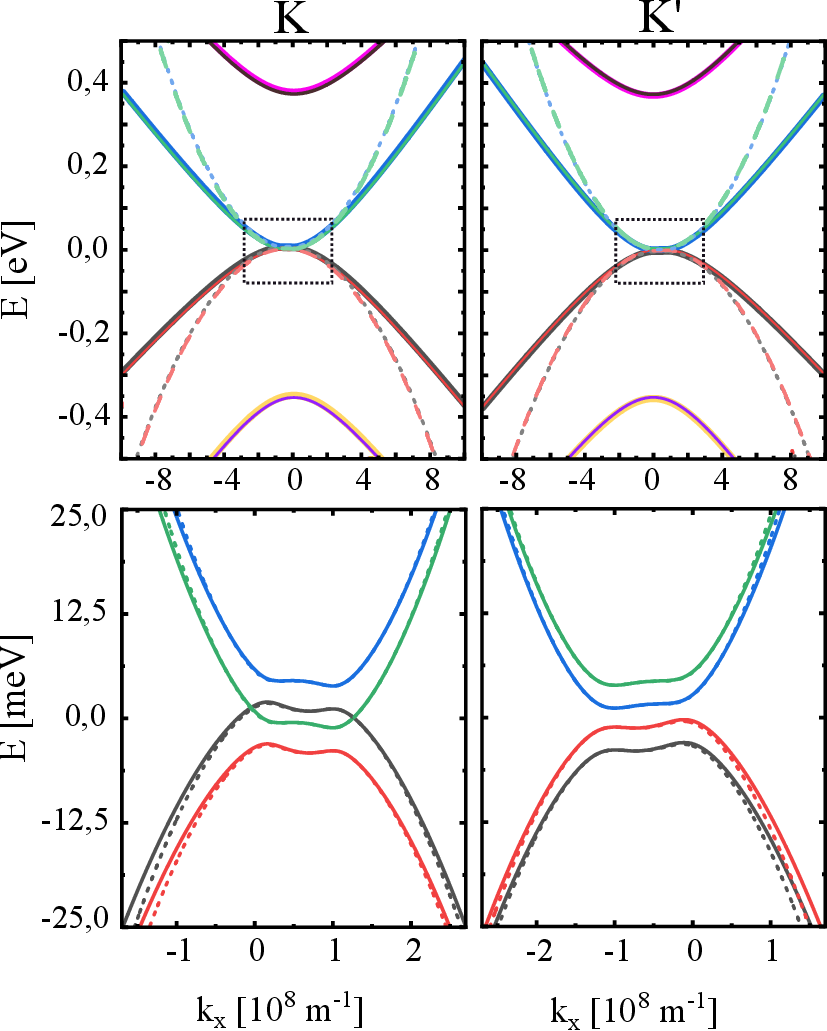}
\caption{Energy spectrum around the K and K' points, obtained based on $\mathbf{k}\cdot\mathbf{p}$ model described by Hamiltonian (\ref{eq:H_kp}) (a,b); and low-energy states around the K and K' points (c,d), where the eigenvalues of $\mathbf{k}\cdot\mathbf{p}$ Hamiltonian (solid lines) are compared with the eigenvalues corresponding to the low-energy reduced Hamiltonian given by Eq.(\ref{eq:HB1A2}) (dotted lines).}
\label{fig:fig2}
\end{figure}


\begin{table}[t!]
    \centering
    \caption{Parameters fitted by Zollner et al.~\cite{Zollner_exsotic} to the model  Hamiltonian~\ref{eq:H_kp}. Numerical data presented in this manuscript are calculated for these parameters. }
    \resizebox{0.7\textwidth}{!}{ 
    \begin{tabular}{cccccc}
    \hline\hline
 Hopping parameters {\small{$\mathrm{[eV]}$}}: &    & $\gamma_{0}$  & $\gamma_{1}$  & $\gamma_{3}$  & $\gamma_{4}$ \\
    \hline
    &    & 2.432 & 0.365 & - 0.273 & - 0.164\\
        \hline\hline
  Proximity-induced SO {\small{$\mathrm{[meV]}$}}: &    & $\lambda_{\scriptscriptstyle{I}}^{A1}$ &  $\lambda_{\scriptscriptstyle{I}}^{B1}$   & $\lambda_{\scriptscriptstyle{I}}^{A2}$ & $\lambda_{\scriptscriptstyle{I}}^{B2}$ \\
         \hline
               &  & 0 & 0 & 1.132\, &\,\,  - 1.132\\
               \hline\hline
   Proximity-induced EX {\small{ $\mathrm{[meV]}$}}: &  & $\lambda_{\scriptscriptstyle{EX}}^{A1}$   & $\lambda_{\scriptscriptstyle{EX}}^{B1}$  & $\lambda_{\scriptscriptstyle{EX}}^{A2}$  & $\lambda_{\scriptscriptstyle{EX}}^{B2}$   \\
         \hline
       &  & - 3.874\, &\, - 3.874 & 0 & 0 \\
          \hline\hline
    Other parameters: &  &\,\,\, \,\,\,$a$ &  & $\Delta$ & \\
    \hline
     & &\,\,\,\,\,\,\,\,\,\,\,\,2.5 & {\small{[$\AA$]}} \,\,\,\,\,\,\,\,& 8.854 & {\small{$\mathrm{[meV]}$}} \\
     \hline
    \end{tabular}}
    \label{tab:tab1}
\end{table}


Figure~\ref{fig:fig2} presents the electronic spectrum around the K and K' points, obtained based on Hamiltonian~(\ref{eq:H_kp}) (solid lines) and calculated for the parameters collected in Tab.~\ref{tab:tab1}. The band structure consists of four pairs of subbands (for each K point): two pairs, form the valence and conduction bands close to the Fermi level, and two other pairs correspond to the valence and conduction bands shifted in energy away from the Fermi level by $\pm\gamma_{1} = \pm 0.36$eV. The latter conduction and valence bands are formed mainly from the $p_z$ orbitals at atoms $A1$ and $B2$, whereas the former bands are formed mainly from the orbitals of atoms $A2$ and $B1$.
In the following considerations we assume $\lambda_{{\scriptscriptstyle{I}}}^{\scriptscriptstyle{A2}} = \lambda_{{\scriptscriptstyle{I}}}$, $\lambda_{{\scriptscriptstyle{I}}}^{\scriptscriptstyle{B2}} = -\lambda_{{\scriptscriptstyle{I}}}$, $\lambda_{{\scriptscriptstyle{EX}}}^{{\scriptscriptstyle{A1}}} = \lambda_{{\scriptscriptstyle{EX}}}^{\scriptscriptstyle{B1}} =\lambda_{{\scriptscriptstyle{EX}}}$

\subsection{Reduced low-energy Hamiltonian}

As the transport properties are related to the four low-energy bands (for each Dirac point) in the vicinity of the Fermi level, one can further reduce the $8\times8$ Hamiltonian~(\ref{eq:H_kp}) to a simpler $4\times4$ form. We derived the reduced Hamiltonian using the Green function method~\cite{McCann2006,Abergel2010}. In the first step the Hamiltonian~(\ref{eq:H_kp}) was written in the basis (B1$\uparrow$,{B1$\downarrow$},{A2$\uparrow$},{A2$\downarrow$}, {A1$\uparrow$},{A1$\downarrow$},{B2$\uparrow$}, {B2$\downarrow$}) and separated into $2\times2$ blocks:
\begin{equation}
\label{eq:H_blocks}
	\hat{H}^{\tau}=
	\begin{pmatrix}
		\mathbb{H}_{11} & \mathbb{H}_{12} \\
		\mathbb{H}_{21} & \mathbb{H}_{22}  \\
	\end{pmatrix}
\end{equation}
In consequence one can define the block $\mathbb{H}_{22}$ that is formed by A1-B2 dimer (related to the higher energy states), and the block $\mathbb{H}_{11}$ that is formed by low-energy states. Next, the Hamiltonian~(\ref{eq:H_blocks}) has been expanded with respect to the parameter $p = 1/\gamma_{1}$. This procedure allows one effectively to exclude the atomic sites involved in the A1-B2 dimer bond. Accordingly, the Green function related to (\ref{eq:H_blocks}) has the form:
\begin{equation}
\label{eq:G_blocks}
	G=
	\begin{pmatrix}
		\mathbb{G}_{11} & \mathbb{G}_{12} \\
		\mathbb{G}_{21} & \mathbb{G}_{22}  \\
	\end{pmatrix},
\end{equation}
where $G_{11}$ contains information on the low-energy states and can be used to determine the effective reduced Hamiltonian~\cite{McCann2006}. Using the definition of the Green function one can write  
\begin{eqnarray}
\label{eq:G_def}
	G & = & 
	\begin{pmatrix}
		\mathbb{H}_{11} - \epsilon  & \mathbb{H}_{12} \\
		\mathbb{H}_{21} & \mathbb{H}_{22} - \epsilon  \\
	\end{pmatrix} ^{-1} = \begin{pmatrix}
		\mathbb{G}_{11}^{0-1}  & \mathbb{H}_{12} \\
		\mathbb{H}_{21} & \mathbb{G}_{22}^{0-1}   \\
	\end{pmatrix} ^{-1}
\end{eqnarray}
where 
\begin{equation}\label{eq:G11}	
	\mathbb{H}_{\alpha \alpha}^{0}  = (	\mathbb{H}_{\alpha \alpha}^{0}  - \epsilon)^{-1}.
\end{equation}
Evaluation of (\ref{eq:G_def}) gives:
\begin{equation}
\label{eq:G11^-1}
	\mathbb{G}_{11}^{-1} + \varepsilon = \mathbb{H}_{11} - \mathbb{H}_{12}\mathbb{G}_{22}^{0}\mathbb{H}_{21}.
\end{equation}
Assuming that $|\varepsilon| \ll \gamma_{1}$, the expression defining $\mathbb{G}_{22}^{0}$ can be expanded with respect to the parameter  $p = 1/\gamma_{1}$. Finally, the reduced Hamiltonian corresponding to $\mathbb{G}_{11}$ can be written in the form: 
\begin{equation}
\begin{split}
\label{eq:HB1A2}
\hat{H}^{\scriptscriptstyle{\tau}}_{{\scriptscriptstyle{B_{1}A_{2}}}}  & = -\frac{v^2}{\gamma_{1}} \left(1+ \gamma_{40}^{2} \right) \left( (k_x^2 - k_y^2) \hat{\eta}_{x} - 2 k_x k_y \hat{\eta}_{y}\right) \otimes  \hat{s}_{0}\\
         & - 2 \frac{v^2}{\gamma_{1}} \gamma_{40} k^{2} (\hat{\eta}_{0}\otimes \hat{s}_{0})  - v \gamma_{30} (\tau k_{\scriptscriptstyle{x}}\hat{\eta}_{x} + k_{\scriptscriptstyle{y}}\hat{\eta}_{y}) \otimes  \hat{s}_{0}  \\ 
                & - \lambda_{\scriptscriptstyle{EX}} (\hat{\eta}_{+} \otimes \hat{s}_{z}) + \lambda_{\scriptscriptstyle{I}}\tau (\hat{\eta}_{-} \otimes \hat{s}_{z}  )
                + V (\hat{\eta}_{z} \otimes \hat{s}_{0} ),\\ 
               \end{split}    
\end{equation}
where $\gamma_{30} = \gamma_{3}/\gamma_{0}$,  $\gamma_{40} = \gamma_{4}/\gamma_{0}$, and $\hat{\eta}_{\alpha}$,$\hat{\eta}_{0}$  represent Pauli matrices and identity matrix acting in the B1-A2 dimer space, $k_{\scriptscriptstyle{\pm}} = k_{\scriptscriptstyle{x}} \tau \pm ik_{\scriptscriptstyle{y}} $, and $\hat{\eta}_{\pm} = $ $\frac{1}{2}(\sigma_{z} \pm \sigma_{0})$. 

The eigenvalues of Hamiltonian~(\ref{eq:HB1A2}) take the form:
\begin{equation}
\label{eq:E_12_full_kx_ky}
		E_{1,2}^\tau  = - F^{\tau}_{\mathbf{k}} \pm \frac{1}{2}(\lambda_{\scriptscriptstyle{EX}} - \tau \lambda_{\scriptscriptstyle{I}}) - 2 \frac{\gamma_{40}}{\gamma_{1}} v^{2}k^{2} 
\end{equation}
\begin{equation}
\label{eq:E_34_full_kx_ky}
E_{3,4}^\tau  =  F^{\tau}_{\mathbf{k}}
              \pm \frac{1}{2}(\lambda_{\scriptscriptstyle{EX}}-\tau \lambda_{\scriptscriptstyle{I}}) - 2 \frac{\gamma_{40}}{\gamma_{1}} v^{2}k^{2} 
\end{equation}
where
{\small{
\begin{equation}
F^{\tau}_{\mathbf{k}}= \left[ v^{2} \gamma_{30}^{2} k^{2} +\frac{v^{4}}{\gamma_{1}^{2}} (1 + \gamma_{40}^{2})^{2} k^{4} + \frac{1}{4}(2V\pm(\lambda_{\scriptscriptstyle{EX}} + \tau\lambda_{\scriptscriptstyle{I}}))^{2}
             - \frac{\tau}{2} \gamma_{1}\gamma_{3}(1+\gamma_{40}^{2}) k_{x}(k_{x}^{2}-3 k_{y}^{2}) \right]^{1/2} .
\end{equation}
}}
The energy spectrum related to the reduced Hamiltonian is presented in Fig.~\ref{fig:fig2} by the dashed lines. From the bottom panel of Fig.~\ref{fig:fig2} one can see that in the energy window between -25meV and +25meV the reduced model is a good approximation of the $\bf{k}\cdot\bf{p}$ model.

The topological properties of the electronic structure are described by the Berry curvature:
\begin{equation}
\label{eq:BC_def}
\bO_{j} = \nabla_{\bm{k}} \times \bm{A}_{j}(\bm{k}) 
\end{equation}
where $\bm{A}_{j} = i \langle \psi_{j} | \nabla_{\bm{k}} | \psi_{j} \rangle$ is the Berry connection. 

\begin{figure} [t!]
\includegraphics[width=0.55\textwidth]{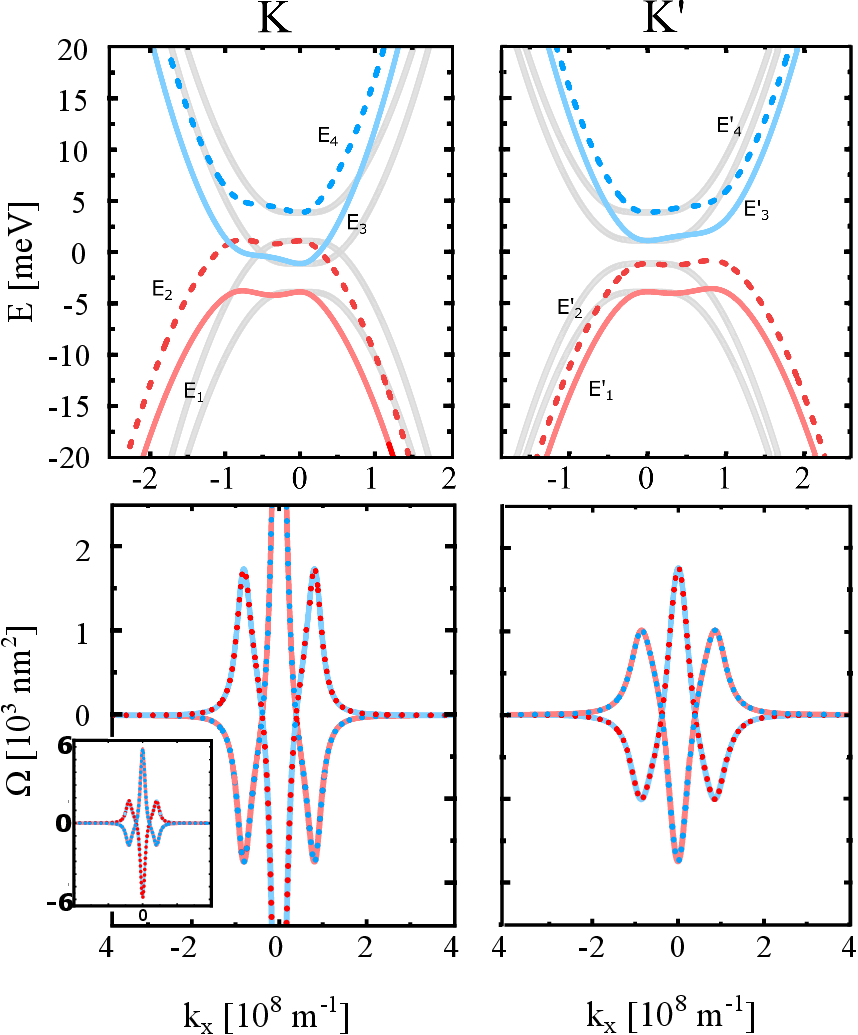}
	\caption {Electronic band structure corresponding to the reduced Hamiltonian (\ref{eq:HB1A2}) describing ex-so-tic vdW structure  (top panel). Red and blue curves present Eigenvalues given by Eqs. (\ref{eq:E_12_full_kx_ky}) and (\ref{eq:E_34_full_kx_ky}). For comparison the band structure for $\gamma_{3,4} = 0$ is plotted in grey. The bottom panel presents the Berry curvature corresponding to the bands presented in the upper panel. }
	\label{fig:fig3}
\end{figure}

Effects of interlayer hoppings between the sites A1-A2 and B1-B2, defined by $\gamma_{4}$, and of the hoppings between the sites A2-B1, defined by $\gamma_{3}$ (see Fig.~\ref{fig:fig1}), are small and give only a correction to the above expressions.
In fact, the hoppings $\gamma_{3,4}$ introduce a small tilting of the energy branches as well as their shift  from the K/K' point, as depicted in Fig.\ref{fig:fig3}, where the eigenvalues of the reduced Hamiltonian are plotted for zero gate voltage (solid gray lines denote the eigenvalues~(\ref{eq:E_12_full_kx_ky}) and (\ref{eq:E_34_full_kx_ky}) for $\gamma_{3,4} = 0$). In the context of transport characteristics studied in this manuscript, the assumption $\gamma_{3,4} = 0$ does not affect general trends but only leads to small quantitative changes. In turn, a big advantage of such a simplification is the possibility of obtaining analytical formulas for transport characteristics.

The eigenvalues of Hamiltonian~(\ref{eq:HB1A2}) for the case ${\gamma_{3,4}=0}$ take the simple forms: 
\begin{equation}
\label{eq:simpE_12}
		E_{1,2}^\tau =  -\sqrt{\frac{v^{4}}{\gamma_{1}^{2}} k^{4} + \left(V\pm\frac{1}{2}(\lambda_{\scriptscriptstyle{EX}} + \tau\lambda_{\scriptscriptstyle{I}})\right)^{2}} \pm \frac{1}{2}(\lambda_{\scriptscriptstyle{EX}} - \tau \lambda_{\scriptscriptstyle{I}}),
\end{equation}

\begin{equation}
 \label{eq:simpE_34}
		E_{3,4}^\tau =  \sqrt{\frac{v^{4}}{\gamma_{1}^{2}} k^{4} +\left(V\pm\frac{1}{2}(\lambda_{\scriptscriptstyle{EX}}
+ \tau\lambda_{\scriptscriptstyle{I}})\right)^{2}} \pm \frac{1}{2}(\lambda_{\scriptscriptstyle{EX}} - \tau\lambda_{\scriptscriptstyle{I}}).
\end{equation}
In turn, 
Berry curvatures for the valence bands $E^{\tau}_{1,2}$ 
take following simple forms:
\begin{equation}
\label{eq:simpBC_12}
\Omega_{1,2}^{\tau} =\tau  \frac{4\pi \frac{v^{4}}{\gamma_{1}^{2}} k^{2} \left(V\pm\frac{1}{2}(\lambda_{\scriptscriptstyle{EX}} + \tau \lambda_{\scriptscriptstyle{I}})\right)}{ \left( \frac{v^{4}}{\gamma_{1}^{2}} k^{4} + \left(V\pm\frac{1}{2}(\lambda_{\scriptscriptstyle{EX}} + \tau\lambda_{\scriptscriptstyle{I}}) \right)^{2} \right)^{3/2}},
\end{equation}
where $\tau = \pm 1$ indicates the valley K/K', respectively, and the sign $\pm$ corresponds to band $E^\tau_{1,2}$, respectively. In turn, Berry curvature for the conduction bands $E^{\tau}_{3,4}$ reads:
\begin{equation}
\label{eq:simpBC_34}
\Omega_{3}^{\tau} = - \Omega_{1}^{\tau} \qquad \Omega_{4}^{\tau} = - \Omega_{2}^{\tau}.
\end{equation}

 \begin{figure} [t!]
\includegraphics[width=0.75\textwidth]{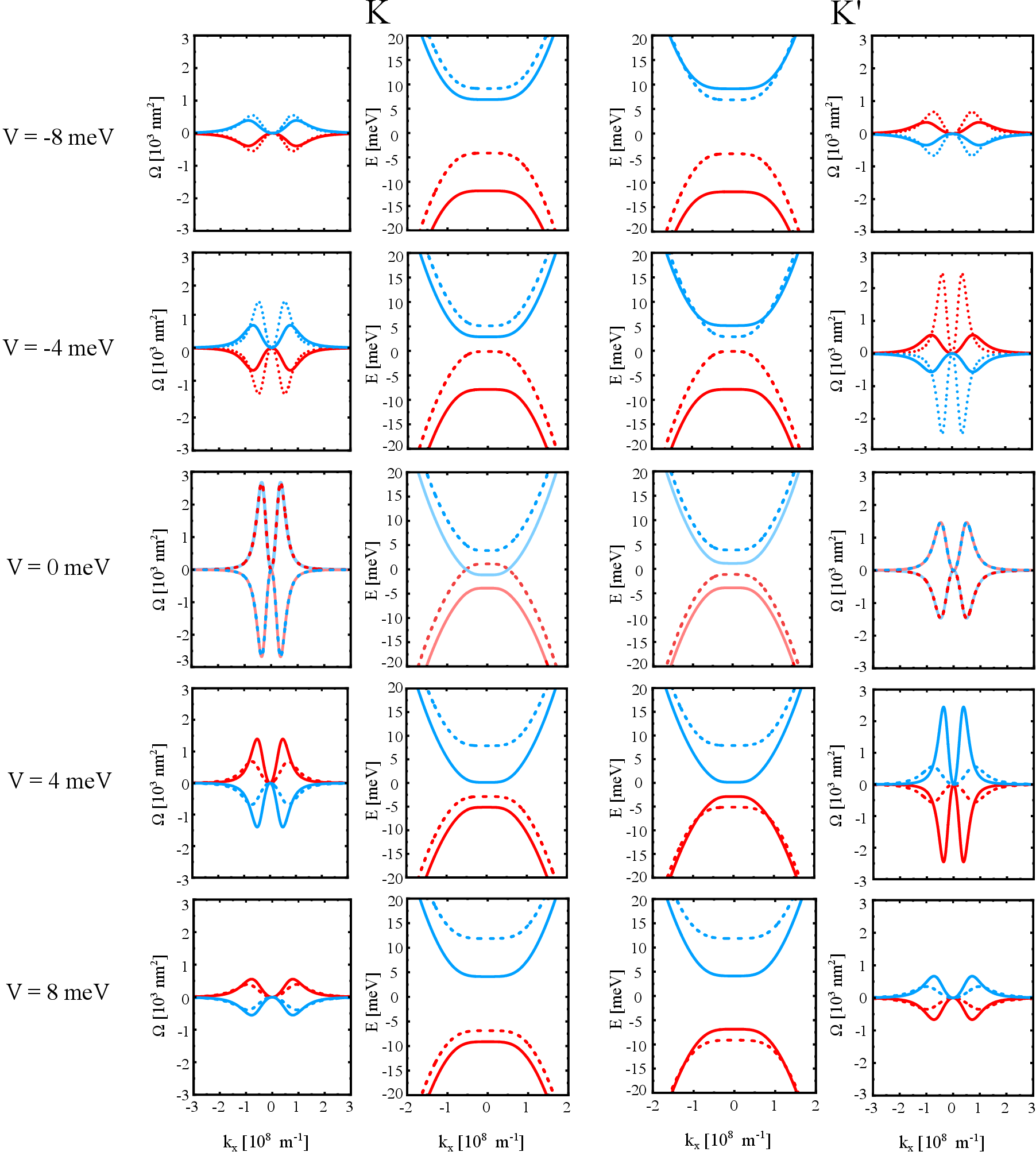}
	\caption {Band structure and the corresponding Berry curvature of ex-so-tic structure at K and K' point for indicated values of the gate voltage. Here, $\gamma_{3,4} = 0$ and the other parameters are listed in Tab.~\ref{tab:tab1}. The dashed and solid lines indicate positive and negative spin polarization, $s_{z} = \pm 1/2$, respectively. }
	\label{fig:fig4}
\end{figure}

Figure~\ref{fig:fig4} shows how the energy dispersion and Berry curvature change with the gate voltage (here defined in the energy units). By changing the gate voltage from -8 meV to +8 meV, one can observe swapping between the domination  of exchange or spin-orbit coupling in the valence and conduction electronic states. For example,  the spin polarizations of the conduction bands in the  K and K' points are identical for ${V=8\mathrm{meV}}$, which suggests the dominant role of exchange interaction in the corresponding electronics states,  whereas the spin polarizations of the valence bands are opposite in the K and K' points, which indicates on the dominant role of spin-orbit interaction in the electronic states. This is, because the spin-orbit interaction does not break the time-reversal symmetry, and thus leads to the opposite spin splitting in the K and K' points, whereas the spin splitting due to the exchange interaction (that leads to breaking of the time-reversal symmetry) should be the same  in the K and K' points. Accordingly, tuning the gate voltage leads to  swapping of the character of  spin splitting in the valence and conduction bands, i.e., to the swapping between spin-orbital and exchange dominated character of the electronic states at the fixed Fermi level. Another important feature of the ex-so-tic structure is the fact that the energy spectrum in the K and K's points is substantially different, which opens the route towards valley-contrasting phenomena in these structures.


\section{Method}
\label{sec:Method}
The transverse dc electric and spin conductivity can be written in terms of Green's function formalism in the following valley-dependent form:
\begin{equation}
\label{eq:sigma_xy}
\sigma_{xy}^{\tau} = \lim_{\omega \to 0} \frac{e^{2} \hbar}{\omega} \int \frac{d \varepsilon}{2\pi} \int \frac{d^{2}\mathbf{k}}{(2\pi)^{2}} \mathrm{Tr} \left[\hat{v}_{x}^{\tau} G^{\tau}_{\mathbf{k}}(\varepsilon + \omega) \hat{v}_{y}^{\tau}G^{\tau}_{\mathbf{k}}(\varepsilon ) \right]
\end{equation}

\begin{equation}
\label{eq:sigma_spin_xy}
\sigma_{xy}^{s_{z}\,\tau} = \lim_{\omega \to 0} \frac{e^{2} \hbar}{\omega} \int \frac{d \varepsilon}{2\pi} \int \frac{d^{2}\mathbf{k}}{(2\pi)^{2}} \mathrm{Tr} \left[\hat{j}_{x}^{s_z\,\tau} G^{\tau}_{\mathbf{k}}(\varepsilon + \omega) \hat{v}_{y}^{\tau}G^{\tau}_{\mathbf{k}}(\varepsilon ) \right]
\end{equation}
where $\hat{v}_{\alpha} = \frac{1}{\hbar} \frac{\partial \hat{H}_{\scriptscriptstyle{B1A2}}^{\tau}}{\partial k_{\alpha}}$ denotes the velocity operator ($\alpha = {x,y}$), and  $\hat{j}_{x}^{s_z\,\tau} = \frac{1}{2}[\hat{v}_{x}^{\tau},\hat{S}_{z}]_{+}$ is the spin current density operator, with the spin operator defined as $\hat{S}_{z} = \frac{\hbar}{2}\hat{\eta}_{0} \otimes \hat{s}_z$. Furthermore, $G_{\mathbf{k}}^{\tau}$ is the casual Green's function defined as $G_{\mathbf{k}}^{\tau} = [(\varepsilon + \mu + i\delta \mathrm{sign}(\varepsilon))\hat{\eta}_{0} \otimes \hat{s}_{0} - \hat{H}_{\scriptscriptstyle{B1A2}}^{\tau}]^{-1}$, where $\mu$ denotes the chemical potential, and $\delta \to 0^{+}$ (as we consider the clean limit). Taking into account contributions from both valleys (K and K') one finds the following expressions for the anomalous and spin Hall conductivity:
\begin{equation}
\sigma_{xy}^{AH} = \sigma_{xy}^{K} + \sigma_{xy}^{K'} :=\mathrm{AHC}
\end{equation}
\begin{equation}
\sigma_{xy}^{SH}  = \sigma_{xy}^{s_z\,K} + \sigma_{xy}^{s_z\,K'} := \mathrm{SHC}
\end{equation}

In the clean limit, when we consider only topological contribution to the Hall conductivity,  Eq.~(\ref{eq:sigma_xy}) can be rewritten in terms of  the Berry curvature as follows:
\begin{equation}
\sigma_{xy}^{\tau} = \frac{e^{2}}{\hbar} \sum_j \int \frac{d^{2} \mathbf{k}}{(2\pi)^{2}} \Omega_{j}^{\tau} f(E_{j})
\end{equation}
where $f(E_{j})$ denotes the Fermi-Dirac distribution function for the j-th subband (for details see e.g. ~\cite{XiaoRevModPhys2010,Dyrdal2017}).

To explore valley contrasting Hall effects, one needs to subtract the transverse charge (spin) Hall conductivity for the K and K' points. Accordingly, we use the following definitions for valley and valley spin Hall effects:
\begin{equation}
\sigma_{xy}^{VH} = \sigma_{xy}^{K} - \sigma_{xy}^{K'} :=\mathrm{VHC}
\end{equation}
\begin{equation}
\sigma_{xy}^{VSH}  = \sigma_{xy}^{s_z\,K} - \sigma_{xy}^{s_z\,K'} := \mathrm{VSHC}
\end{equation}

\section{Results and discussion}
\label{sec:Results}
Here, we present and discuss the results obtained for the anomalous and spin Hall effects, as well as for two valley-contrasting phenomena known as the valley Hall effect and valley spin Hall effect. First, we present the results obtained when neglecting the interlayer hopping integrals $\gamma_{3,4}$. In such a case, all the numerical results presented in the next two subsections have been obtained based on fully analytical solutions (all formulas are presented in the supplementary material). Then, in the last subsection we will discuss the influence of a  nonzero $\gamma_{3,4}$.

\subsection{Anomalous and Spin Hall Effects}

 \begin{figure} [t!]	\includegraphics[width=0.9\textwidth]{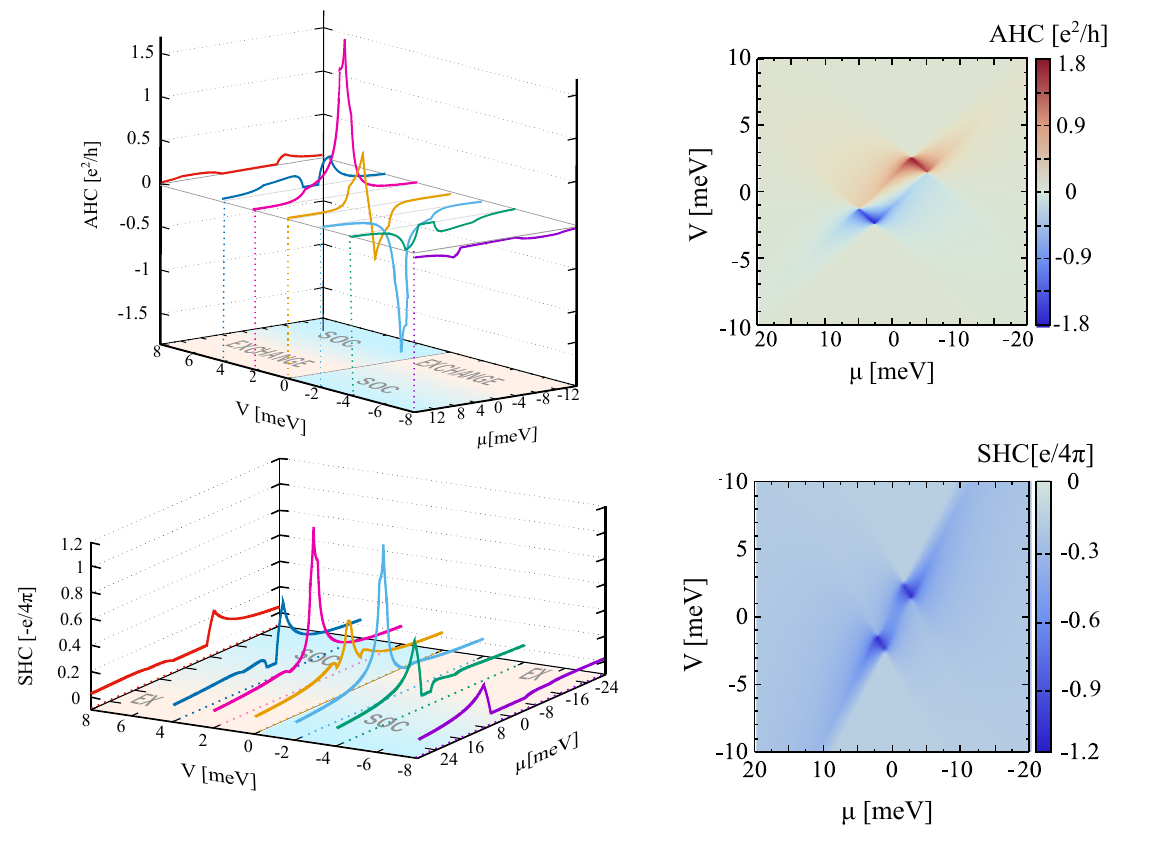}
	\caption { Anomalous Hall conductivity and spin Hall conductivity as a function of chemical potential, $\mu$ and gate voltage,$V$. The other parameters are listed in the Tab.~\ref{tab:tab1}. }
	\label{fig:fig5}
\end{figure}

Figure~\ref{fig:fig5} presents the anomalous Hall and spin Hall conductivities as a function of chemical potential and gate voltage. Both AHE and SHE do not achieve quantized values in the energy gap. Thus, there is no transition to a topologically nontrivial phase. The AHC displays very sharp peaks in a well-defined range of chemical potential and gate voltage. More precisely, AHE is negative for positive chemical potentials, $\mu$, and negative values of gate voltage, $V$, while it is positive  for negative chemical potentials and positive values of gate voltage. Moreover, it displays  very sharp peaks only for a very narrow range of $\mu$ and $V$.  In the top right plot in
Fig.~\ref{fig:fig5}, the peaks in AHC are well-seen as  red and blue hot spots, that appear for the gate voltage range, where  the spin-orbit interaction dominates in the valence or conduction band.
In Fig.~\ref{fig:fig5} one can see, that SHC behaves in a similar way, however, it is positive in the whole range of $\mu$ and $V$, and reveals peaks when the Fermi energy is in the valence or conduction band and the gate voltage ensure  the dominant role of spin-orbit interaction in the corresponding electronic states. 
These characteristics of the anomalous and spin Hall effects can be very attractive in the context of applications and construction of electronic elements with very strong electronic signals only in a very specific energy range.

\subsection{Valley Hall effects}

 \begin{figure} [t!]	\includegraphics[width=0.9\textwidth]{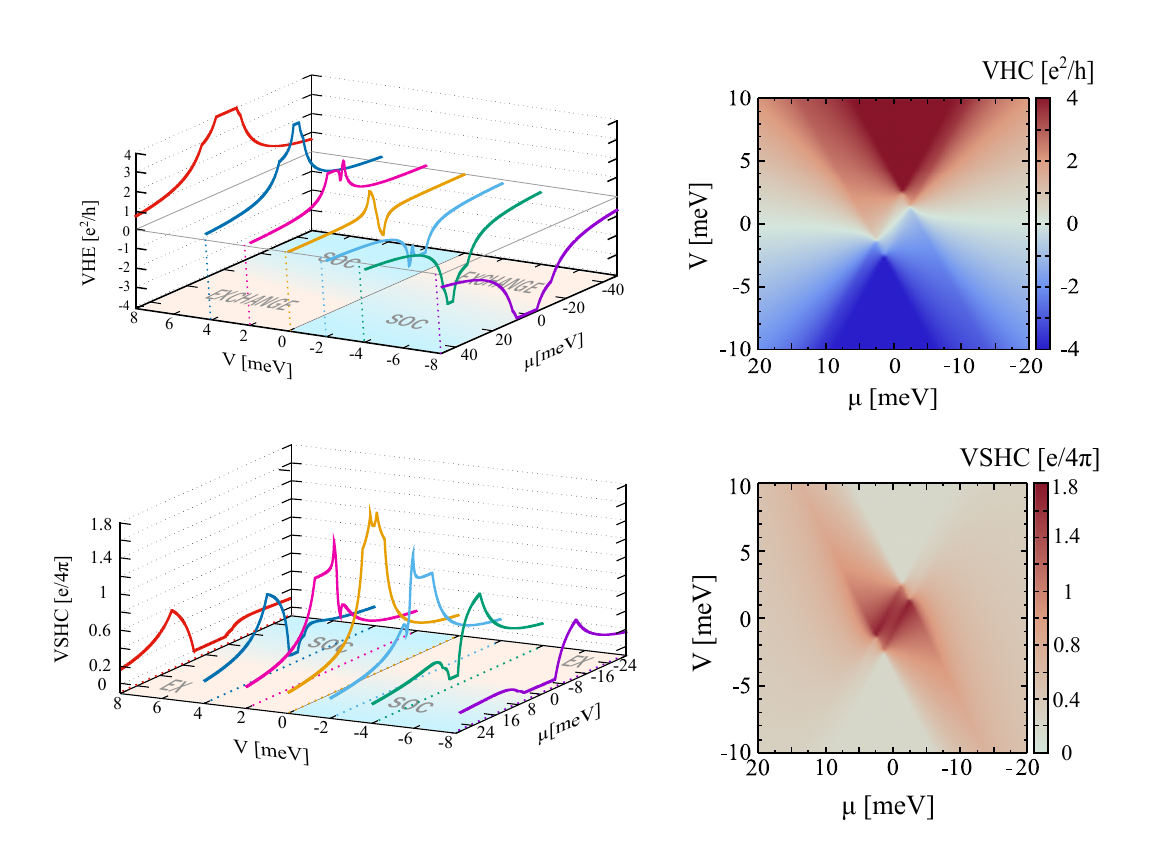}
	\caption { Valley Hall conductivity and valley spin Hall conductivity as a function of chemical potential, $\mu$ and gate voltage,$V$. The other parameters are listed in the Tab.~\ref{tab:tab1}. }
	\label{fig:fig6}
\end{figure}

An important feature of the  ex-so-tic structure under consideration is a clear distinction between the energy dispersion around the K and K's points. This allows one to measure the system response that comes from only a single valley or at least is dominated by electronic states from a specific valley. Recently, the valley Hall effects in graphene-based systems have been considered in Refs~\cite{Dyrdal2017,Islam2016,Tarucha_NatPhys2015}. However, as long as the energy bands are not distinguishable in the K and K' points, the valley-dependent transport properties are very difficult to be measured. The ex-so-tic structure seems to be a big step forward in the development of valleytronics.   

Figure~\ref{fig:fig6} shows behaviour of the 
valley Hall and valley spin Hall conductivities as a function of chemical potential and gate voltage.
The valley Hall effect is nonzero in the system under consideration, and VHC reaches the quantized value when the Fermi energy lies inside the energy gap. This quantized value changes from +4$\mathrm{e}^{2}/h$ to $- 4\mathrm{e}^{2}/h$, depending on the gate voltage. It is worth noting that the quantized valley Hall effect appears in the system because the Hall conductivity associated with electronic states in the K and K' points achieves a quantized value equal $\pm2\mathrm{e}^{2}/h$ respectively for K/K' point, when the Fermi level is inside the energy gap. This explains not only the quantized VHC but also a zero AHC when the Fermi energy lies in the energy gap. 
Figure~\ref{fig:fig6} also  shows, that in ex-so-tic structures one can expect a nonzero valley spin Hall effect. The corresponding  VSHC, in contrast to the SHC, is positive for the whole range of parameters $V$ and $\mu$. Moreover, the well-defined picks in VSHC are observed for the Fermi levels in the valence or conduction bands, depending on the value of gate voltage that ensures the dominant role of exchange interaction in electronic states. 

\subsection{Effect of interlayer shopping $\gamma_{3}$ and $\gamma_{4}$}

\begin{figure} [t!]	\includegraphics[width=0.87\textwidth]{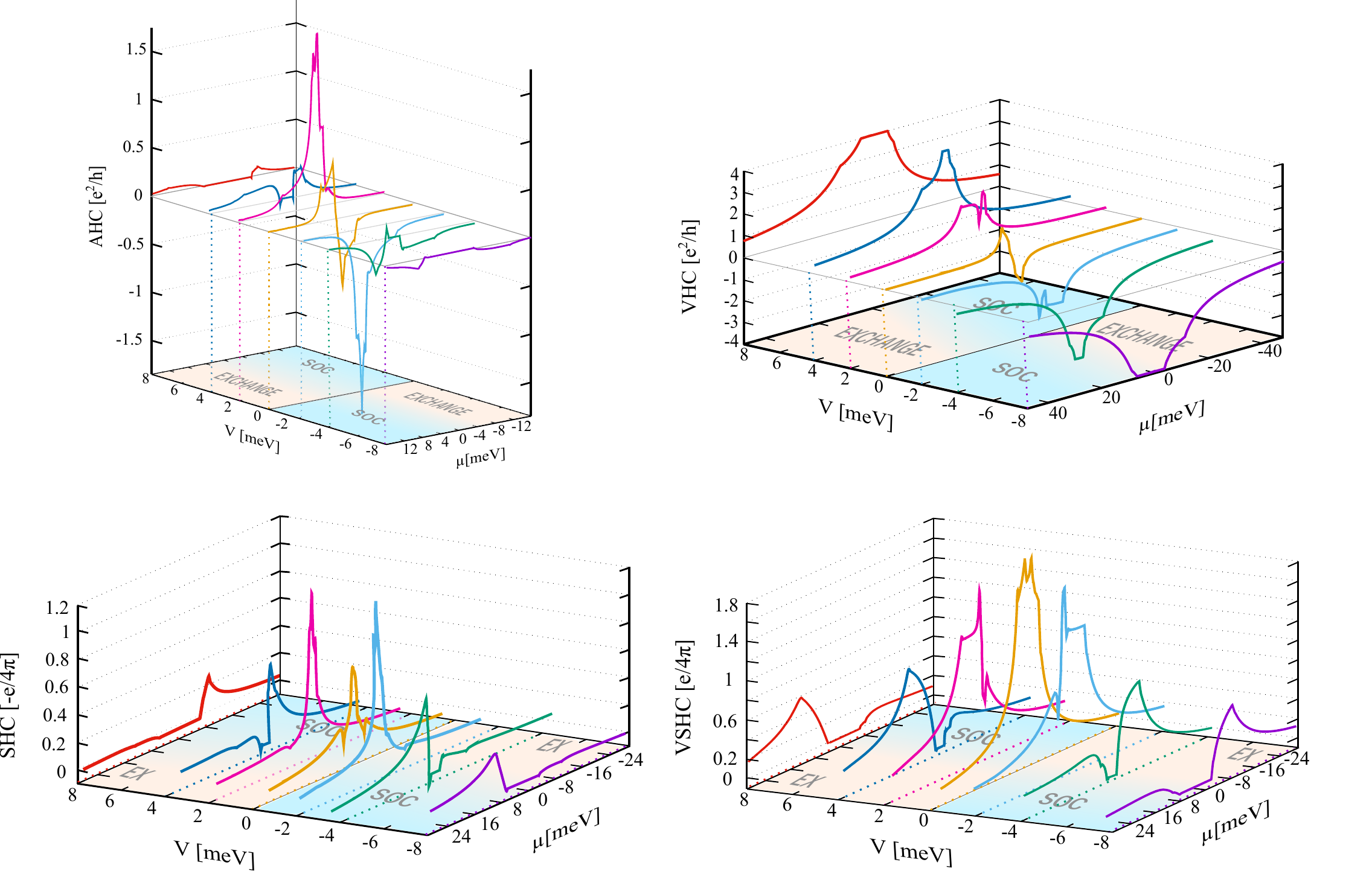}
	\caption {The anomalous, spin, valley and valley spin Hall conductivities as a function of chemical potential and gate voltage in case of nonzero $\gamma_{3,4}$ hoppings. The other parameters are listed in the Tab.~\ref{tab:tab1}. }
	\label{fig:fig7}
\end{figure}

\begin{figure} [t!]	\includegraphics[width=0.8\textwidth]{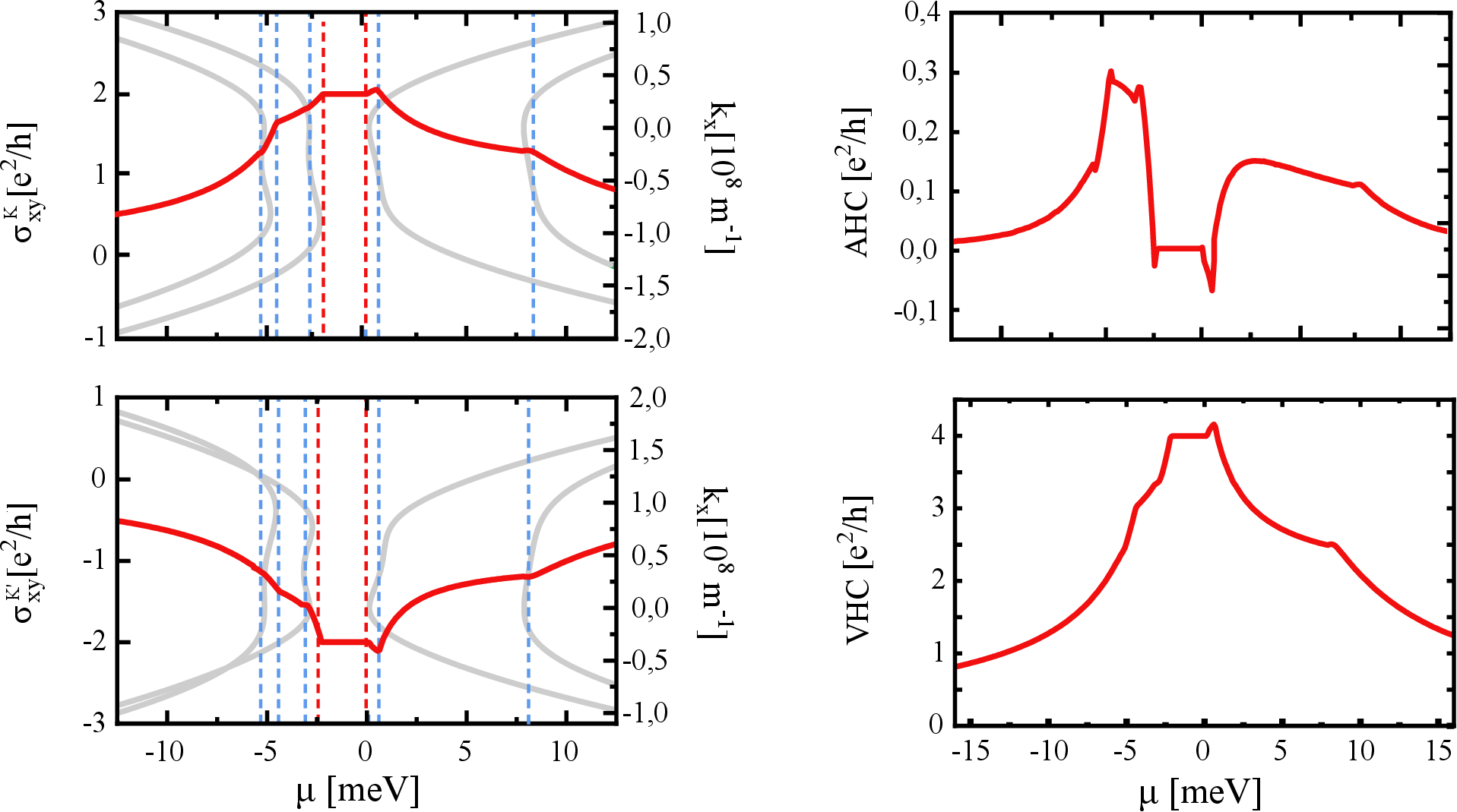}
	\caption {Charge Hall conductivity in K and K' points (left panel) as a function of chemical potential and anomalous and valley Hall effect (right panel) as a function of chemical potential in case of nonzero $\gamma_{3,4}$ hoppings. The other parameters are listed in the Tab.~\ref{tab:tab1}. }
	\label{fig:fig8}
\end{figure}

Now, we discuss the effect of nonzero hopping integrals $\gamma_{3,4}$. Fig.~\ref{fig:fig7} presents all the  Hall conductivities discussed in the previous two subsections, but with $\gamma_{3,4}$ taken into account. The results have been obtained numerically and do not differ qualitatively from those presented in Figs.~\ref{fig:fig5} and~\ref{fig:fig6}. However, in all these characteristics, one can identify additional kinks or spikes. These sharp spikes reflect the valley-contrasting physics and band structure. This is clearly visible in Fig.~\ref{fig:fig8}, where in the left panel we present charge Hall conductivity for the K and K' points as a function of chemical potential, as well as  their sum (that is the AHC) and their difference (that is the VHC), see the right panel. In this plot, it is clearly seen that each kink or spike in the Hall conductivity reflects the position of the local extremes in energy bands.


\section{Conclusions}
\label{sec:Conclusions}

We investigated the anomalous, spin, and valley-contrasting effects in an ex-so-tic van-der-Waals structure consisting of a bilayer graphene deposited on a ferromagnetic 2D insulator, such as CGT, and covered by a semiconducting 2D crystal, such as WS$_2$. Accordingly, one 2D crystal ensures proximity-induced exchange coupling, and the second one ensures proximity-induced spin-orbit coupling, and both affect the electronic structure of BLG. 
We have derived the reduced Hamiltonian describing the low-energy spectrum of the ex-so-tic structure, and subsequently we have used the Green's function formalism to calculate  the specific charge and spin Hall conductivities. Interestingly, for a fixed position of Fermi level, one can tune the gate voltage to obtain strong nonzero anomalous or spin Hall conductivity. The well-picked AHC (or SHC) characteristic can be useful for new elements for spintronics (such as diodes or transistors). Moreover, we showed that the system can be tuned to the specific range of gate voltage and chemical potentials, for which it displays the quantized valley Hall conductivity. The quantized VHC can be equal $\pm 4 \mathrm{e}^{2}/h$ depending on the sign of gate voltage. 

\section*{Acknowledgement}
This work has been supported by the Norwegian Financial Mechanism under the Polish-Norwegian Research Project NCN GRIEG '2Dtronics', project no. 2019/34/H/ST3/00515.

\section*{References}
\bibliographystyle{iopart-num}
\bibliography{bib.bib}

\section*{Data availability statement}
All data that support the findings of this study are included within the article (and any supplementary files).

\end{document}